\documentstyle[nato,numreferences,epsf,graphicx,psfrag,times,amsmath,mathptm,cite]{crckapb}
\begin{opening}
\title{Solitons and spontaneous symmetry breaking in 2 and 4 dimensions}
\author{Manfried Faber, Joachim Wabnig}
\institute{Atominstitut der \"osterreichischen Universit\"aten,\\
           TU Wien, Wiedner Hauptstr. 8-10, A--1040 Vienna, Austria}
\author{Andrei N. Ivanov}
\institute{State Technical University, Department of Nuclear Physics, 
195251
St. Petersburg, Russian Federation}
\runningtitle{Solitons and spontaneous symmetry breaking}
\runningauthor{M.\ Faber, A.\ Ivanov}
\end{opening}
\begin{document}
%
%
\begin{abstract}
  We show that mass generation in 1+1 and 3+1 dimensions may occur
  together with spontaneous symmetry breaking.
\end{abstract}
%
%
%
\renewcommand{\thefootnote}{\fnsymbol{footnote}}
\footnotetext[0]{Presented by Manfried Faber
at the NATO Advanced Research Workshop ``Confinement, Topology,
and other Non-Perturbative Aspects of QCD'',
January 21--27, 2002, Star\'a Lesn\'a (Slovakia).  
Supported in part by Fonds zur F\"orderung der
Wissenschaftlichen Forschung P13997-TPH.}
\renewcommand{\thefootnote}{\arabic{footnote}}

\section{Thirring model in the chirally broken phase}

Skyrme's conjectured \cite{Sky58} that the solitons of the sine-Gordon
model have the properties of fermions and couple by an interaction of
the Thirring model type. The sine-Gordon model is a model of a bosonic
field $\vartheta(x)$ in 1+1--dimensional space--time with a Lagrangean
which in Coleman's \cite{Col75} notation reads
\begin{eqnarray}\label{LsG}
{\cal L}(x) =
\frac{1}{2}\partial_{\mu}\vartheta(x)\partial^{\mu}\vartheta(x) +
\frac{\alpha}{\beta^2}\,(\cos\beta\vartheta(x) - 1).
\end{eqnarray}
The Thirring model \cite{Thi58} describes a self-coupled Dirac-field
$\psi(x)$ in 1+1 dimension
\begin{eqnarray}\label{LThi}
{\cal L}(x) = \bar{\psi}(x)(i\gamma^{\mu}\partial_{\mu} - m)\psi(x) -
\frac{1}{2}\,g\,\bar{\psi}(x)\gamma^{\mu}\psi(x)\bar{\psi}(x)
\gamma_{\mu}\psi(x),
\end{eqnarray}
where $m$ is the mass of the fermion field and $g$ is a dimensionless
coupling constant. The field $\psi(x)$ is a spinor field with two
components $\psi_1(x)$ and $\psi_2(x)$. The $\gamma$--matrices are
defined in terms of the well--known $2\times 2$ Pauli matrices,
$\gamma^{\;0} = \sigma_1$, $\gamma^1 = - i\sigma_2$, $\gamma^5 =
\gamma^0\gamma^1 = -i\sigma_1\sigma_2 = \sigma_3$.

Coleman suggested a perturbative approach to the understanding of the
equivalence between the sine--Gordon and the Thirring model. He
developed a perturbation theory with respect to $\alpha$ and $m$ in
order to compare the $n$--point Green functions in the sine-Gordon and the
massive Thirring model in coordinate representation. Under the
assumption of the existence of these two theories, Coleman concluded
that they should be equivalent if the coupling constants $\beta$ and
$g$ obey the relation \cite{Col75}
\begin{eqnarray}\label{ColRel}
\frac{4\pi}{\beta^2} = 1 + \frac{g}{\pi}
\end{eqnarray}
and the operators $\psi(x)$ and
$\vartheta(x)$ satisfy the Abelian bosonisation rules
\begin{eqnarray}\label{BosRules}
Z\,m\,\bar{\psi}(x)\Bigg(\frac{1\mp \gamma^5}{2}\Bigg)\psi(x) =
-\frac{\alpha}{\beta^2}\,\,e^{\textstyle \pm i\beta \vartheta(x)},
\end{eqnarray}
where the constant $Z$ depends on the regularisation \cite{Col75}.

The common point of all approaches to the solution of the massless
Thirring model \cite{Kla68}--\cite{FM88} and to the derivation of the
equivalence between the sine-Gordon and the massive Thirring model
\cite{Col75,FM88,DNS92} is a quantisation of the fermionic system
around the trivial perturbative vacuum.

\begin{figure}
\centerline{\scalebox{1.0}{\includegraphics{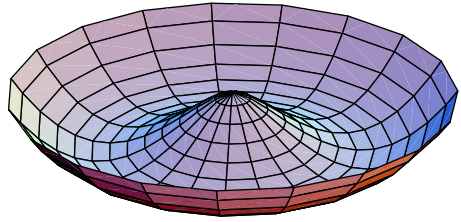}}}
\caption{The energy density ${\cal E}$ of 
Eq.(\ref{EneDens}) as a function of $M$ and $\alpha_{\rm A}$ 
for $2\pi/g = 2{\rm ln}\,2$.}
\label{fig2}
\end{figure}

The Thirring model coincides with the Nambu--Jona--Lasinio (NJL) model
in 1+1--dimensional space--time. It is well--known that the NJL model
is a relativistic covariant generalisation of the BCS theory \cite{FI01} of
superconductivity. Within operator and path integral formalism one can show \cite{FI01} for the massless Thirring model that the energy density is described by
\begin{eqnarray}\label{MEq0Ene}
{\cal E}(M) = \frac{1}{4\pi}\,\Bigg[M^2\,{\rm ln}\frac{M^2}{\Lambda^2}
- (\Lambda^2 + M^2)\,{\rm ln}\Bigg(1 + \frac{M^2}{\Lambda^2}\Bigg) +
\frac{2\pi}{g}\,M^2\Bigg],
\end{eqnarray}
where $\Lambda$ defines the interaction region and $M$ is the mass of the quasi--particles. The energy density of the perturbative vacuum is therefore maximal and the state of minimal energy is characterised by the non-trivial solution of the gap--equation
\begin{eqnarray}\label{GapEqu}
M = \frac{\Lambda}{\displaystyle \sqrt{e^{\textstyle
2\pi/g} - 1}}.
\end{eqnarray}
Under chiral rotations of the massless Thirring fermion fields
\begin{eqnarray}\label{ChiRot}
\psi(x,t) \to {\psi\,}'(x,t) &=& e^{\textstyle i\gamma^5\alpha_{\rm
A}}\psi(x,t),\nonumber\\ \bar{\psi}(x,t) \to {\bar{\psi}\,}'(x,t) &=&
\bar{\psi}(x,t)\,e^{\textstyle i\gamma^5\alpha_{\rm A}}.
\end{eqnarray}
the wave function of the non--perturbative vacuum transforms
\begin{eqnarray}\label{TraWF}
&|\Omega\rangle \to |\Omega; \alpha_{\rm A}\rangle
= \prod_{\textstyle p^1}\Big[u_{\textstyle p^1}
+ v_{\textstyle p^1}\,e^{\textstyle -2i\alpha_{\rm A}\varepsilon(p^1)}\,
a^{\dagger}(p^1)b^{\dagger}(-p^1)\Big]|0\rangle,\\
&u_{\textstyle p^1} = \sqrt{\frac{1}{2}\Bigg( 1 +
{\displaystyle \frac{|p^1|}{\sqrt{(p^1)^2 + M^2}}}\Bigg)}\;,\; v_{\textstyle
p^1} = \varepsilon(p^1)\,\sqrt{\frac{1}{2}\Bigg( 1 -
{\displaystyle \frac{|p^1|}{\sqrt{(p^1)^2 + M^2}}}\Bigg)},
\end{eqnarray}
but the energy density ${\cal E}(M)$ is invariant, see Fig.~\ref{fig2}.

Following the dynamics of the chiral phase $\alpha_{\rm A}$ and the identification $\alpha_{\rm A}=\beta\vartheta$ leads to the Lagrangean \cite{FI01}
\begin{eqnarray}\label{EneDens}
{\cal L}_{\rm eff}(x) = \frac{\beta^2}{16\pi}\,\Big(1 - e^{\textstyle
-2\pi/g}\,\Big)\,\partial_{\mu}\vartheta(x)\,\partial^{\mu}\vartheta(x),
\end{eqnarray}
and a relation between the coupling constants $\beta$ and $g$
\begin{eqnarray}\label{CouRel}
\frac{8\pi}{\beta^2} = 1 - e^{\textstyle -2\pi/g}
\end{eqnarray}
differing from Coleman's relation (\ref{ColRel}).

\begin{figure}
\centerline{\scalebox{1.0}{\includegraphics{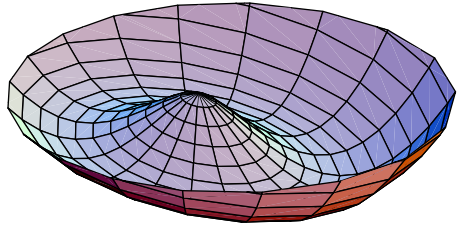}}}
\caption{The energy density ${\cal E}$ in the massive Thirring model as a function of $M$ and $\alpha_{\rm A}$ for $2\pi/g = {\rm ln}\,2^2$ and $4\pi m/g = 0.2$ in units of$\Lambda$.}
\label{fig3}
\end{figure}
For the massive Thirring model the chiral symmetry of the energy
density is explicitly broken, see Fig.~\ref{fig3}, and the effective
Lagrangian gets the form of the Lagrangian of the sine-Gordon model
(\ref{LsG}) with the relation
\begin{eqnarray}\label{alphaRel}
\alpha =\beta^2\,\frac{m M}{g} =
-\,m\,\beta^2\,\langle\bar{\psi}\psi\rangle + \frac{m^2}{g}\,\beta^2.
\end{eqnarray}
It is interesting to note that Eq.~(\ref{alphaRel}) leads to a
relation for the mass of the sine-Gordon soliton $M_{\rm sol} =
8\sqrt{\alpha}/\beta^2$ resembling the Gell-- Mann--Oakes--Renner
low--energy theorem for the mass spectrum of low--lying pseudoscalar
mesons \cite{GOR68}. For $\beta^2 > 8$ the mass $\sqrt{\alpha}$ of the
sine-Gordon bosons becomes greater than the mass of solitons.  This
implies that in the bosonised version of the massive Thirring model
with $\beta^2 \ge 8\pi$, according to Eq.~(\ref{CouRel}) the creation
of non--perturbative solitons is energetically preferable with respect
to the creation of bosons. The bosons are decoupled from the system
and there exist practically only solitons.

The topological current of the sine-Gordon model coincides with the
Noether current of the massive Thirring model related to the $U_{\rm
  V}(1)$ invariance \cite{FI01}, responsible for the conservation of
the fermion number in the massive Thirring model, the topological
charge of the sine-Gordon model has the meaning of the fermion number.
Since many--soliton solutions obey Pauli's exclusion principle, this
should prove Skyrme's statement \cite{Sky58} that the sine-Gordon
model solitons can be interpreted as massive fermions.  Thus, via
spontaneously broken chiral symmetry the massive Thirring fermions get
converted into extended particles with the properties of fermions and
masses much heavier than their initial mass.

\section{Stable solitons in 3+1 dimensions}

We will now show that spontaneous symmetry breaking in four dimensions
may also lead to stable solitons of finite mass, to extended
particles, which at large distances behave like dual Dirac monopoles,
like electrical charged particles \cite{Fab99} whose interaction
originates in their topology.

\begin{figure}
  \centering
  \includegraphics[width=0.6\textwidth]{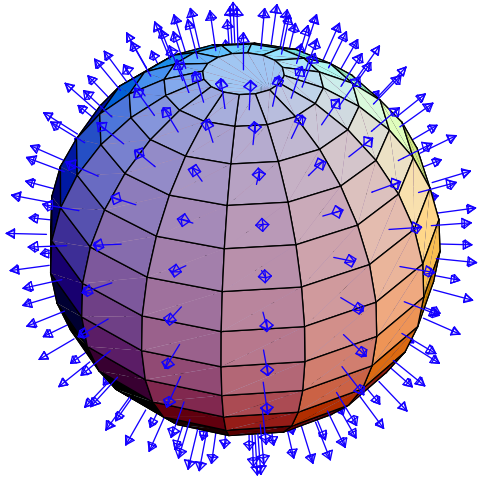}
  \caption{Dirac monopole as a static hedgehog of an $\mathrm{S}^2$ field $\vec{n}(x)$ in physical space $\mathrm{R}^3$}
  \label{colouredhedgehog}
\end{figure}

The Dirac string can be dissolved by a transition from a description with a dual U(1) gauge field
\begin{equation}\label{DualPot}
C_\varphi = - \frac{e_0}{4\pi \varepsilon_0} \frac{\Gamma_\phi}{r \sin \vartheta} = \frac{e_0}{4\pi \varepsilon_0} \frac{1 - \cos \vartheta}{r \sin \vartheta}, C_t =  C_r = C_\vartheta = 0
\end{equation}
to a field of unit vectors $\vec{n}(x)$ in the form of a hedgehog
\begin{equation}\label{hedgehog}
\vec{n} = \frac{\vec{r}}{r}, \quad \vec{n}^{\,2}=1,
\end{equation}
see Fig.~\ref{colouredhedgehog}. This $\vec{n}$-field has a singularity at the origin and two degrees of freedom which can be identified with the two degrees of freedom of the electro-magnetic field in the vacuum. Since no magnetic monopoles have been found until now we prefer the interpretation of Dirac monopoles as electrically charged particles.

The $\vec{n}$ representation of the Dirac monopole is equivalent to the original description in terms of the gauge field \cite{Fab99} and makes it evident how one can dissolve the singularity at the origin. Removing the condition $\vec{n}^{\;2}=1$ by multiplying  $\vec{n}$ with a factor $0 \le \sin \alpha \le 1$ and reducing the length of the vectors in the centre to zero ($\alpha=0$) the hedgehog field can be regularised. This corresponds to a transition to a field of unimodular quaternions, to an SU(2)-field
\begin{equation}
Q = q_0 + i\vec{\sigma}\,\vec{q} = e^{\textstyle i\alpha \vec{\sigma}\,
\vec{n}}, \quad \vec{n}^{\;2} = 1, \quad 0 \le \alpha \le \pi,
\end{equation}
where the fields $\alpha, \vec{n}, q_0$ and $\vec{q}$ depend on the
position $x^\mu=(\vec{r},t)$ in Minkowski space-time and
$\vec{\sigma}$ are the Pauli matrices.\footnote{We use the summation
convention that any capital Latin index that is repeated in a product
is automatically summed on from 1 to 3. The arrows on variables in the
internal ``colour'' space indicate the set of 3 elements
$\vec{q}=(q_1, q_2, q_3)$ or $\vec{\sigma}=(\sigma_1, \sigma_2,
\sigma_3)$ and $\vec{q}\,\vec{\sigma}= q_K \sigma_K$.}

We can now introduce the Lagrangian of dual SU(2)-QCD by a geometrical concept. As suggested by gauge-theories, we would like to interpret the field strength as internal curvature in colour space, in the parameter space $\mathrm{S}^{\;3}$ of SU(2). This can be done via the definition of a connection $\vec{\Gamma}_\mu(x)$ between local internal coordinate systems. Using $i \sigma_K Q$ as basis vectors of the tangential space at $Q$ we define the connection field as the tangential vector $\vec{\Gamma}^\mu$ from $Q(x)$ to $Q(x+dx^\mu)$
\begin{equation}\label{dualgaugefield}
\vec{C}^\mu = - \frac{e_0}{4 \pi \varepsilon_0} \vec{\Gamma}^\mu, \quad \partial^\mu Q = i \vec{\Gamma}^\mu \vec{\sigma} Q.
\end{equation}
The relation between the dual gauge field and the connection is implied by the transition to the SI--system. The connection field $\vec{C}^\mu$ looks only superficially like a pure gauge field which would lead to the rather unnatural definition of curvature zero for the unit sphere $\mathrm{S}^{\;3}$ and to field strength zero. The above definition (\ref{dualgaugefield}) differs from the pure gauge definition by a factor of $2$.

The electric flux through a rectangle in configuration space is given
by the corresponding area in the tangential plane of
$\mathrm{S}^{\;3}$. This suggests to identify the dual field strength
tensor with the curvature tensor $\vec{R}^{\mu \nu}$
\begin{equation}\label{dualfieldstrength}
\hspace{0.8mm}{^*}\hspace{-0.8mm}\vec{F}^{\mu \nu} = - \frac{e_0}{4 \pi \varepsilon_0 c} \vec{R}^{\mu \nu}, \quad \vec{R}^{\mu \nu} = \vec{\Gamma}^\mu \times \vec{\Gamma}^\nu.
\end{equation}
The connection field $\vec{\Gamma}^\mu$ obeys the Maurer-Cartan equation\cite{Fab99}
\begin{equation}
\vec{\Gamma}^\mu \times \vec{\Gamma}^\nu = \frac{1}{2} \left( \partial^\nu \vec{\Gamma}^\mu - \partial^\mu \vec{\Gamma}^\nu \right).
\end{equation}
The curvature tensor can therefore be represented in a form which is well known from non-abelian gauge theories
\begin{equation}
\vec{R}^{\mu \nu} = \partial^\nu \vec{\Gamma}^\mu - \partial^\mu \vec{\Gamma}^\nu - \vec{\Gamma}^\mu \times \vec{\Gamma}^\nu.
\end{equation}
In those theories the gauge field $\vec{\Gamma}^\mu$ is the basic field and can't be derived from a soliton field.

It is well--known from QCD that a Lagrangean which contains the square
of the curvature only dissolves monopoles. Therefore we supplement the
Lagrangean of dual QCD by a compressing Higgs-potential, a function of
tr$Q=2q_{\;0}$ with the minimum at $q_{\,0}=0$ and monotonically
increasing with $q_0^2$ allowing for spontaneous symmetry breaking
\begin{equation}\label{Lagrangean}
{\cal L} =  - \frac{\alpha_f \hbar c}{4 \pi}\left( \frac{1}{4} \; \vec{R}_{\mu \nu} \vec{R}^{\mu \nu} + \Lambda(q_0) \right) = - \frac{1}{4\mu_0} \; \hspace{0.8mm}{^*}\hspace{-0.8mm}\vec{F}_{\mu \nu} \hspace{0.8mm}{^*}\hspace{-0.8mm}\vec{F}^{\mu \nu}  - \frac{\alpha_f \hbar c}{4 \pi r_0^4} \overset{\circ}{\Lambda}(q_0),
\end{equation}
with Sommerfeld's fine--structure constant $\alpha_f = \frac{e_0^2}{4 \pi \varepsilon_0 \hbar c }$.

Minimising the the energy for the hedgehog-ansatz $\vec{n}=\vec{r}/r, \alpha=\alpha(r), \alpha(0)=0, \alpha(\infty)=\pi/2$
leads to a non--linear differential equation for the profile function $q_0(\rho)$ in the natural radial coordinate $\rho=\frac{r}{r_0}$
\begin{equation}\label{nlDE}
\partial^2_\rho q_0 \; + \; \frac{(1 - q_0^2) q_0}{\rho^2} \; - \; \frac{1}{2} \rho^2 \partial_{q_0}\overset{\circ}{\Lambda}(q_0) \; = \; 0.
\end{equation}

As long as the above mentioned general requirements for the potential are fulfilled the special choice of the potential does influence the properties of the soliton core only, i.e.\ the high energy properties. For the potential $\overset{\circ}{\Lambda}(q_0)=\frac{7}{2}(q_0)^{4}$ there exists an analytical solution
\begin{equation}\label{solution}
q_0(\rho) = \frac{1}{1+\rho^2},
\end{equation}
which results in the energy for a single soliton
\begin{equation}
E_1 \; = \; \frac{\alpha_f \hbar c}{r_0} \frac{7\pi}{16} \quad \text{with} \quad \alpha_f \hbar c = 1.44 \text{~MeV~fm}.
\end{equation}
Comparing this expression with the rest energy of the electron of 0.511 MeV, one gets $r_0 = 3.87$ fm, a value close to the classical electron radius of $r_e=\alpha_f\frac{\hbar}{m_e c}=2.82$ fm.
There are three contributions to the radial energy density, the expression in the square braces in
\begin{equation}\label{staticE}
H \; = \; \frac{\alpha_f \hbar c}{r_0} \int_0^\infty{d\rho} \left[ \frac{(1 - q_0^2)^2}{2 \rho^2} + (\partial_\rho q_0)^2  + \rho^2 \overset{\circ}{\Lambda}(q_0) \right],
\end{equation}
two contributions from the electric field and one from the potential energy. The contribution of the colour component of the electric field in the direction of $\vec{n}$, the ``radial field'' decreases with $1/r^2$ and has the largest range. At large distances it provides the classical Coulomb field of a point--like charge. It reaches its maximum at $r_0$ and decreases in the centre due to the shrinking of the spherical volume. The nasty divergence of the classical Coulomb field is obviously removed. The ``tangential'' part of the electric energy density and the potential energy density decay with $\rho^6$ and are concentrated in the region below $r_0$. By topological reasons for all shapes of he potential $\Lambda(q_0)$ the radial field must behave for large distances like $1/r^2$ and the state of the vacuum depends on the direction from the centre of the soliton.

A restriction of the fields to $q_0=0$, to those at the minimum of the
potential $\Lambda(q_0)$, leads to Maxwell's electrodynamics in a dual
formulation \cite{BBZ94}. In this ``Electrodynamic limit'' the soliton
field degenerates to $Q(x)=i\,\vec{\sigma}\,\vec{n}(x)$ and can be
characterised by the space-time dependence of an
$\mathrm{S}^{\,2}$--field, the unit vector field $\vec{n}$. The two
degrees of freedom of $\vec{n}$ provide the two polarisations of the
free electromagnetic field. Connection field and curvature get very
simple expressions, $\vec{\Gamma}_\mu(x)=\partial^\mu \vec{n} \times
\vec{n}$ and $\vec{R}^{\mu\nu}=\partial^\mu \vec{n} \times
\partial^\nu \vec{n}$, and the gauge symmetry of electrodynamics has a
very intuitive geometrical interpretation. The vectors $i\,\sigma_K\,
Q(x)$, in the tangential plane of the equatorial
$\mathrm{S}^{\,2}_{equ}$ attributed to $x$, define a local coordinate
system for the connection fields $\vec{\Gamma}_\mu(x)$ at $x$. A
$U(1)$ rotation of this local coordinate system around $\vec{n}$
corresponds to the well known $U(1)$ gauge symmetry of Maxwell's
electrodynamics. By a further rotation of these $\vec{n}$ vectors in
$3$ direction \cite{Fab99} one arrives at the dual formulation of
electrodynamics, discussed in \cite{BBZ94}. This transition to the
Abelian description is like combing a sphere and attaches line-like
singularities to the point--like singularities of the hedgehogs, one
gets the well--known Dirac--strings \cite{Dir48}.

We can conclude that also in 3--1 dimensions a model with spontaneous
symmetry breaking may lead to solitons of finite mass. These solitons
have quantised charges following from their topology and
particle--antiparticle annihilation, already at the classical level.
With only three degrees of freedom this model describes the phenomena
of the free Maxwell field, charged particles and their interaction via
the Maxwell field. The only difference to Maxwell's electrodynamics is
in the fact that the charges in Maxwell's electrodynamics can be
arbitrary. In the mentioned model only multiples of the elementary
charge are allowed and this should not be a disadvantage as it agrees
with experiment. This holds also true for magnetic monopoles, objects
that are unstable in this model.


\end{document}